\RequirePackage{lineno}
\documentclass[a4paper,11pt]{article}
\pdfoutput=1
\usepackage{jcappub} 
\usepackage{subcaption}
\usepackage{amsmath}
\usepackage{xspace} 
\usepackage{placeins}
\usepackage{tabulary}
\usepackage{hyperref}
\usepackage[toc,page]{appendix}
\usepackage[T1]{fontenc} 

\newcommand{\PhiPP}{\ensuremath{\Phi_{\rm PP}}\xspace}
\newcommand{\sigmav}{\ensuremath{\langle \sigma v \rangle}\xspace}
\newcommand{\Jfactor}{\ensuremath{\mathrm{J- factor}}\xspace}

\newcommand{\ie}{i.e.\xspace}

\begin{document}

\title{\boldmath Analyzing  $\gamma$ rays of the Galactic Center with Deep Learning }
\author[a,b]{Sascha Caron}
\author[c]{Germ\'an A. G\'omez-Vargas}
\author[a]{Luc Hendriks}
\author[d]{Roberto Ruiz de Austri}

\affiliation[a]{Institute for Mathematics, Astrophysics and Particle Physics, Faculty of Science, Mailbox 79,
Radboud University Nijmegen, P.O. Box 9010, NL-6500 GL Nijmegen, The Netherlands}
\affiliation[b]{Nikhef, Science Park, Amsterdam, The Netherlands}
\affiliation[c]{Instituto de Astrof\'isica, Pontificia Universidad Cat\'olica de Chile, Avenida Vicu\~na Mackenna 4860, Santiago, Chile}
\affiliation[d]{Instituto de F\'isica Corpuscular, IFIC-UV/CSIC, Valencia, Spain}

\emailAdd{scaron@cern.ch}
\emailAdd{ggomezv@uc.cl}
\emailAdd{luc@luchendriks.com}
\emailAdd{rruiz@ific.uv.es}

\abstract{


We present the application of convolutional neural networks to a particular problem in gamma ray astronomy.  Explicitly, we use this method to investigate the origin of an excess emission of GeV $\gamma$ rays in the direction of the Galactic Center, reported by several groups by analyzing {\it Fermi}-LAT data.  Interpretations of this
excess include $\gamma$ rays created by the annihilation of dark matter particles and $\gamma$ rays originating from a collection of unresolved point sources, such as millisecond pulsars. We train and test convolutional neural networks with simulated {\it Fermi}-LAT images based on point and diffuse emission models of the Galactic Center tuned to measured $\gamma$ ray data.
Our new method allows precise measurements of the contribution and properties of an unresolved population of $\gamma$ ray point sources in the interstellar diffuse emission model. The current model predicts the fraction of unresolved point sources with an error of up to 10\% and this is expected to decrease with future work.

}
\newcommand{\germancomment}[1]{\ignorespaces {\large \color{blue} GG: #1}}
\newcommand{\luccomment}[1]{\ignorespaces {\large \color{red} LUC: #1}}

\maketitle
\flushbottom 
\section{Introduction}
\label{sec:introduction}

The {\it Fermi} Large Area Telescope (LAT) has been observing the $\gamma$ ray sky for energies $\gtrsim 100$ MeV since August 2008 with unprecedented angular resolution and sensitivity \cite{2009ApJ...697.1071A}. The $\gamma$ rays produced by the interaction of cosmic ray (CR) particles with matter and radiation fields in the interstellar medium (ISM) of the Milky Way dominate the emission detected with the {\it Fermi}-LAT in the Galactic plane~\cite{Ackermann:2012kna}.  Physical modeling codes, such as GALPROP\footnote{http://galprop.stanford.edu}, calculate the propagation of CRs in the ISM and compute diffuse $\gamma$ ray emission in the same framework \cite{Moskalenko:1997gh,Strong:1998fr}. Each GALPROP run using specific, realistic astrophysical inputs will create a different interstellar emission model (IEM) in the form of a series of templates, each for a particular physical emission mechanism. The usual method to compare these IEMs with data is the so-called { \it template fitting method}~\cite{1996ApJ...461..396M}. A large set of IEMs were compared to {\it Fermi}-LAT data in~\cite{Ackermann:2012pya}, finding that the IEMs reproduce general features of the interstellar $\gamma$ ray emission over the whole sky. However, the template models do not capture all the information contained in the {\it Fermi}-LAT data. The present analysis focuses on an unaccounted component detected in the region of the inner Galaxy, the so-called Galactic center Excess (GCE), using the template fitting method \cite{Goodenough:2009gk, Vitale:2009hr, Hooper:2010mq, Gordon:2013vta, Hooper:2011ti, Daylan:2014rsa, 2011PhLB..705..165B, Calore:2014xka, Abazajian:2014fta,Zhou:2014lva}. 

The {\it Fermi}-LAT collaboration investigated the GCE properties (spectral shape, magnitude, and morphology) by using 6.5 years of observations in Refs. \cite{TheFermi-LAT:2015kwa, 1704.03910}, with the conclusion that the GCE is present in the data despite many sources of uncertainties, but its specific properties are significantly model dependent. The gray band of figure \ref{fig:GCE} presents the set of spectral energy distributions of the GCE found in \cite{1704.03910}, derived from the different systematics sources investigated.  This emission has been interpreted as a possible signal for the annihilation of dark matter (DM) particles or as a signal of unresolved point sources~\cite{Caron:2015wda,Abazajian:2014fta,Mirabal:2013rba,PhysRevD.95.103005}. DM-like signatures observed in other regions of the Galactic plane, where such signals are not expected (control regions), are used to quantify the magnitude of systematic uncertainties due to diffuse emission modeling in the Galactic center and prevent the interpretation of the GCE as a signal of DM \cite{1704.03910}.

Many groups have investigated the possibility that a population of unresolved $\gamma$ ray pulsars could be the origin for the GCE emission, see for instance \cite{McCann:2014dea,Mirabal:2013rba,2015arXiv150402477O,2014JHEAp...3....1Y,Petrovic:2014xra,Cholis:2014lta,Macias:2016nev}. The Galactic bulge is expected to host a population of pulsars due to the high level of past star formation activity in that region \cite{Macquart:2015jfa}. In the whole sky, more than 200 pulsars have been identified in the $\gamma$ ray band\footnote{https://confluence.slac.stanford.edu/x/5Jl6Bg}. Furthermore, using novel statistical methods the authors of \cite{2016PhRvL.116e1103L}, \cite{2016PhRvL.116e1102B} and \cite{Storm:2017arh} claim evidence for the existence of an unresolved population of $\gamma$ ray sources in the inner $20^{\circ}$ of the Galaxy with spatial distribution and collective flux compatible with the GCE.  Recently the {\it Fermi}-LAT collaboration also investigated the pulsar interpretation of the GCE \cite{Fermi-LAT:2017yoi}. Performing a new point source search using 7.5 years of Pass 8 {\it Fermi}-LAT data in a $40^{\circ} \times  40^{\circ}$ box around the Galactic center, they confirm the findings of \cite{2016PhRvL.116e1103L} and \cite{2016PhRvL.116e1102B}. However, some arguments \cite{Hooper:2015jlu,Hooper:2016rap,Haggard:2017lyq,Eckner:2017oul,Bartels:2017vsx} have been raised against this interpretation of the GCE. These works point out that a hypothetical $\gamma$ ray millisecond pulsar (MSP) population in the Galactic bulge with similar properties of extant populations, as in globular clusters and the local Galactic disc, is not able to reproduce the total GCE emission. However, these argument are based on the assumption that an MSP population in the Galactic bulge share similarities with MSP populations elsewhere \cite{Ploeg:2017vai}. Because there are no MSPs detected in the Galactic bulge yet, this assumption might not be true. Therefore, the debate on the nature of the GCE is not yet closed, leaving the possibility of having all or a fraction of the GCE due to DM annihilation or other diffuse sources.

The image texture of the GCE can help to unveil its nature; to determine if the GCE is genuinely diffuse or has a granular morphology \cite{2016PhRvL.116e1103L,2016PhRvL.116e1102B,Storm:2017arh}. A completely diffuse GCE would favor the DM interpretation, whereas cumulative emission of point sources too weak to be detected individually would produce a granular morphology\footnote{If the point sources are too dim or too close to each other, it is tough to distinguish them from diffuse radiation. Also, Ref. \cite{Agrawal:2017pnb} argues that if some fraction of the DM has dissipative interactions, it can form dense DM clumps resulting in a population of $\gamma$ ray point sources in the Galactic center region.}. The {\it template fitting} approach is not sensitive to this difference. This work aims to present a new method based on convolutional neural networks (ConvNets) to determine whether the morphology of the GCE is diffuse or granular. We apply our approach to the data collected with the {\it Fermi}-LAT in the region of the Galactic center, and we use simulations of this area to train and validate the network. The present paper is a proof-of-principle work, and we simplify the models for simulations to a level in which the results on the GCE nature provide valuable information for a more sophisticated implementation of the method in a future publication.

We organize the paper as follows: 
We start with an introduction to the GCE. Section \ref{sec:ConvNet} presents the basics on ConvNets. This technology needs large amounts of data to work, and we use realistic simulations for both training and testing the networks; section \ref{sec:analysis_setup} expands on the setup to create these images. Section \ref{sec:ConvNetsetup} presents the ConvNet designed to make predictions on the fraction of granularity present in the GCE, i.e., how much of the GCE is due to a population of unresolved point sources or a truly diffuse source. Results are presented in section \ref{sec:results}, conclusions and foreseen applications are discussed in Section \ref{sec:conclusion}.

\section{Convolutional Neural Networks}\label{sec:ConvNet}

ConvNets are a class of deep neural networks (DNNs) that are designed to make predictions on visual data \cite{lecun_bottou_bengio_haffner_1998}. Neural networks are used in many areas of research to solve classification or regression queries and are well suited for high dimensional problems (for example Ref. \cite{George:2017fbn}). The authors of ref. \cite{Hezaveh:2017sht} employed a similar methodology for analyzing gravitational lensing. In general, a ConvNet takes an N-dimensional input, transforms it using many linear and nonlinear operations, and produces an M-dimensional output (for example, the information introduced can be an image and the output a probability that the image belongs to a particular class) \cite{NIPS2012_4824}. The result of a ConvNet can be a prediction that a specific input belongs to a specific type (e.g., object detection) or a projection of a regression problem (e.g., this work). The input of a computer vision task is typically an image; a $(w, h, c)$-tensor can be a way to represent the data. The $w$ and $h$ in the tensor denote the width and height of the network and $c$ the number of channels, that can be a color filter or energy band. In this work the number of channels is 1: A single energy-integrated counts map of the Galactic Center region, with a specific fraction of the GCE due to a population of unresolved point sources, free to vary between 0 and 1. We call this fraction the $f_{\rm src}$ parameter. $f_{\rm src}=0$ means that the GCE is composed of only a diffuse source and $f_{\rm src}=1$ implies that the GCE is formed only of the cumulative emission of point sources too dim to be detected individually.

In this work, the ConvNet is trained to predict $f_{\rm src}$ from Galactic center images created with counts in the energy band 1-6 GeV, as this is the band where the GCE is most relevant, see figure~\ref{fig:GCE}~\cite{TheFermi-LAT:2015kwa, 1704.03910}. In Appendix \ref{appendix:convnets} we provide a more thorough introduction to ConvNets, in particular training them (Appendix \ref{appendix:training}).

\section{Images for training and testing the ConvNets}
\label{sec:analysis_setup}

Typically a large dataset is needed to train a ConvNet because they can have up to millions (or billions) of tunable weights (see Appendix \ref{appendix:convnets}). To have an extensive training dataset is a challenge in applying ConvNet technology to the {\it Fermi}-LAT Galactic center data, as there is only one $\gamma$ ray image of this region (e.g., compared to many different pictures of galaxies for galaxy classification). However, the advanced knowledge of the {\it Fermi}-LAT instrument~\cite{2009ApJ...697.1071A} and the availability of state-of-the-art interstellar diffuse models~\cite{Ackermann:2012pya} make it possible to create realistic mock {\it Fermi}-LAT data for training and testing examples.  The goal of the trained ConvNet is to predict $f_{\rm src}$ regardless of flux and location of the unresolved point sources as well as the IEM. To achieve this goal, we create a set of {\it 1.2 million mock training images} of the Galactic center region based on models fitted to real data\footnote{ For training; we create 1.000 realizations of 1.200 different values of $f_{\rm src}$  (1.2 million images). Three samples of 20.000 other $f_{\rm src}$ values were used to test the ConvNets (60.000 images).}.  To determine the optimal amount of pictures in the training set, we train a ConvNet multiple times using different training dataset sizes and compare the accuracy of the $f_{\rm src}$ predictions.

The IEM is a crucial component to generate different realizations of the {\it Fermi}-LAT observation of the Galactic center region. As there is no unique model, we must implement different IEM versions. We use five of the eight alternatives IEMs created to study the systematic effects on supernova remnant (SNR) properties caused by the modeling of the diffuse emission around them~\cite{dePalma:2015tfa}. The models selected represent the largest variation in the input parameters used the set of eight IEMs. The starting point is to create models using GALPROP, with a variation on the three input parameters that according to Ref.~\cite{Ackermann:2012kna} are the most relevant in modeling the Galactic plane diffuse emission\footnote{CR source distribution, the height of the CR propagation halo, and spin temperature of the molecular gas.}. The output of GALPROP is in the form of separated templates for emission associated with gas (atomic and molecular) in four Galactocentric rings and a single model for the inverse Compton emission. The second step in the creation of the IEMs is to tune individual scaling factors for these different template components using all-sky likelihood fit to  {\it Fermi}-LAT data. Finally, we re-fit the five IEMs selected leaving free only the scaling factors of the gas related emission in the first innermost rings. In table \ref{tab:diffuse} we present our selection and use of the five IEMs in this work. We decided to use the models created in reference~\cite{dePalma:2015tfa} as they have more freedom in fitting the diffuse emission in the Galactic plane than other models, as the standard IEM of the {\it Fermi}-LAT team for point-source analysis~\cite{Acero:2016qlg}. We also included in the model the sources listed in the 3FGL catalog \cite{Acero:2015hja}. Finally, we add to the model of the Galactic center area the integrated $\gamma$ ray flux at Earth, $\phi_s$, expected from dark matter annihilation in density distribution, $\rho(r)$, given by

\begin{equation}
\begin{aligned}
   \phi_s(\Delta\Omega) = 
   & \underbrace{ \frac{1}{4\pi} \frac{\sigmav}{2m_{DM}^{2}}\int^{E_{\rm max}}_{E_{\rm min}}\frac{\text{d}N_{\gamma}}{\text{d}E_{\gamma}}\text{d}E_{\gamma}}_{\PhiPP}
   & \times
   \underbrace{\vphantom{\int_{E_{min}}} \int_{\Delta\Omega}\Big\{\int_{\rm l.o.s.}\rho^{2}(r)\text{d}l\Big\}\text{d}\Omega '}_{\Jfactor}\,. 
\end{aligned}
\label{eqn:annihilation}
\end{equation}

Here, the \PhiPP term depends on the particle physics properties of dark matter---\ie, the thermally-averaged annihilation cross section, \sigmav, the particle mass, $m_{DM}$, and the differential $\gamma$ ray yield per annihilation, $\text{d}N_\gamma/\text{d}E_\gamma$, integrated over the experimental energy range from $E_{\rm min}$ to $E_{\rm max}$.
The \Jfactor is the line-of-sight integral through the dark matter distribution integrated over a solid angle, $\Delta\Omega$. 
Qualitatively, the \Jfactor encapsulates the spatial distribution of the dark matter signal, while \PhiPP sets its spectral character. Note that we assume annihilation of Majorana particles.

We use the gNFW density distribution to model the spatial DM distribution \cite{NFW:1997,Navarro:1995iw}:
\begin{equation}\label{eq:gNFW}
\rho(r)=\rho_s\frac{r_s^3}{r^{\gamma} (1 + r_s)^{3-\gamma}}.
\end{equation}
Where $r_s$ is the scale radius (20 kpc) and $\rho_s$ a scale density fixed by the requirement that the local DM density at Galactocentric radius of 8.5 kpc is 0.4 GeV cm$^{-3}$. To model \PhiPP, we follow the DM interpretation of the GCE considered in ref. \cite{Calore:2014xka}, and use a $m_{DM}=$ 50 GeV, and the differential $\gamma$ ray yield from WIMP annihilating into  $b\overline{b}$ final state.

\begin{figure}[t]
 \begin{center}
    \epsfig{file=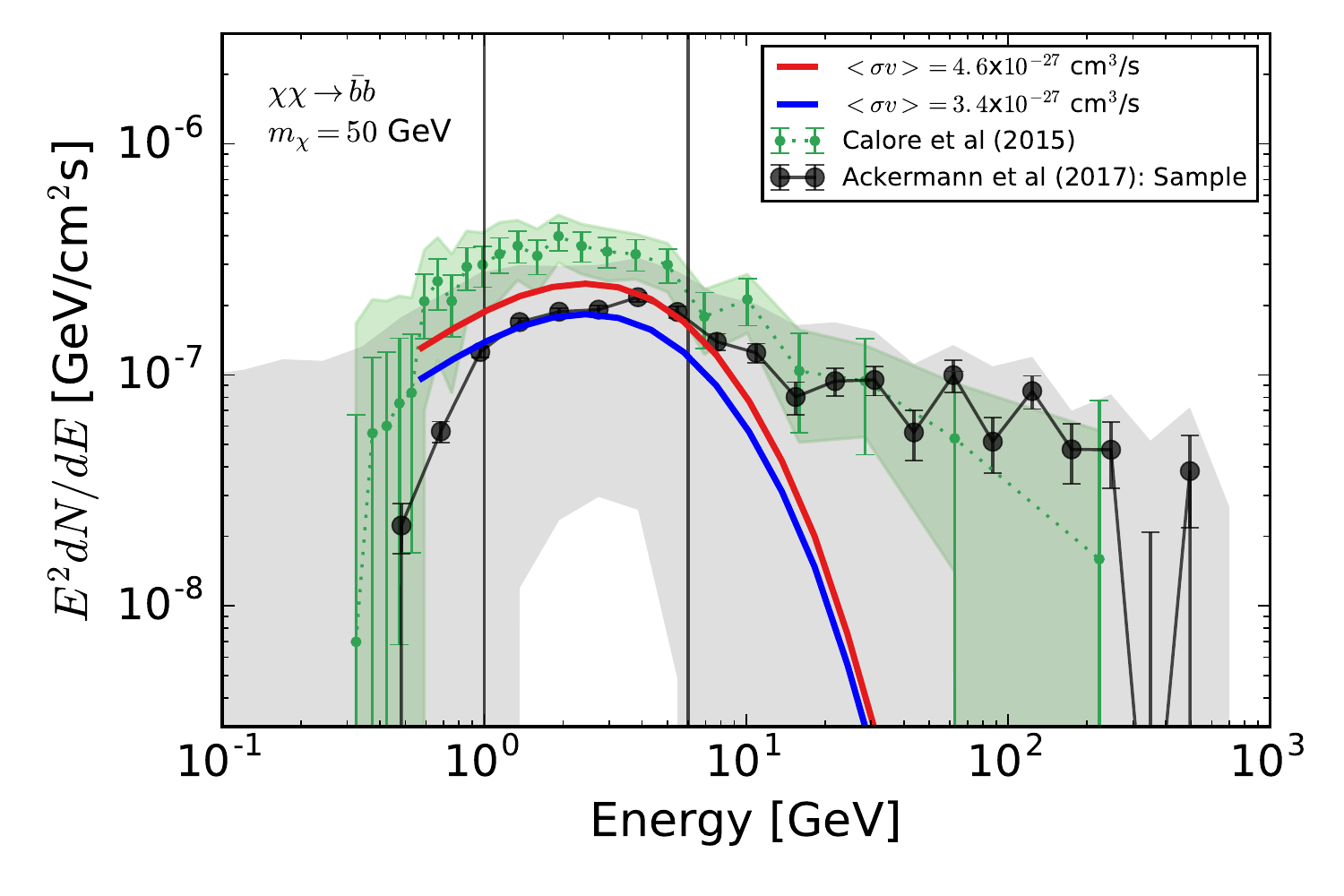} 
 \end{center}
 \caption{Spectral energy distribution of the GCE. In blue and red the spectra from fits performed in this work to the {\it Fermi}-LAT photon data using the models Training A and B in table \ref{tab:diffuse}, respectively. Following \cite{1704.03910} we select the GCE derived using the Sample Model as representative of the GCE spectrum (black points). The gray shaded area represents the variation in the GCE spectra obtained in \cite{1704.03910}. The vertical black lines point out the energy range where the GCE is always present; we use this band to create all ConvNets analyses images. For comparison the GCE spectrum as found in \cite{Calore:2014xka} (green points) is plotted together with the diagonal of the covariance matrix derived there (light green band).}
  \label{fig:GCE}
\end{figure}


\subsection{Template fitting procedure}
\label{sec:fit_procedure_results}

In this work, we use the {\it Fermi}-LAT data collected between 2008 August 4 and 2015 August 2 ({\it Fermi} Mission Elapse Time 239559568 s - 460166404 s). To avoid bias due to residual backgrounds in all-sky or large scale analysis, we select the events belonging to the Pass 8 UltraCleanVeto class and use the corresponding {\tt P8R2\_ULTRACLEANVETO\_V6} instrument response functions; they have the highest purity against contamination from charged particles misclassified as $\gamma$ rays. Furthermore, to reduce the contamination of $\gamma$ rays from interactions of CRs in the upper Earth's atmosphere, we select events with a maximum zenith angle of $100^{\circ}$. We bin the events into 21 energy bins from 500 MeV to 100 GeV for fitting the IEMs to the real $\gamma$ ray sky (note that for the ConvNets we use a single energy bin from 1 to 6 GeV). We only use the events converted in the front part of the {\it Fermi}-LAT, as this is a good compromise between angular resolution and statistics to produce sharp images. Thus we use the same data in fitting and for ConvNets. For the fitting procedure we use a spatial resolution of $0.1^{\circ}$, while for ConvNets we use $0.25^{\circ}$, this is to reduce the Poissonian noise and the computing time of generating images. We re-fit the five IEMs listed in table \ref{tab:diffuse}, including a gNFW template, to {\it Fermi}-LAT data in the inner $15^{\circ}\times 15^{\circ}$ about the Galactic center. 

\begin{figure}[t!]
  \centering
  \includegraphics[width=\textwidth]{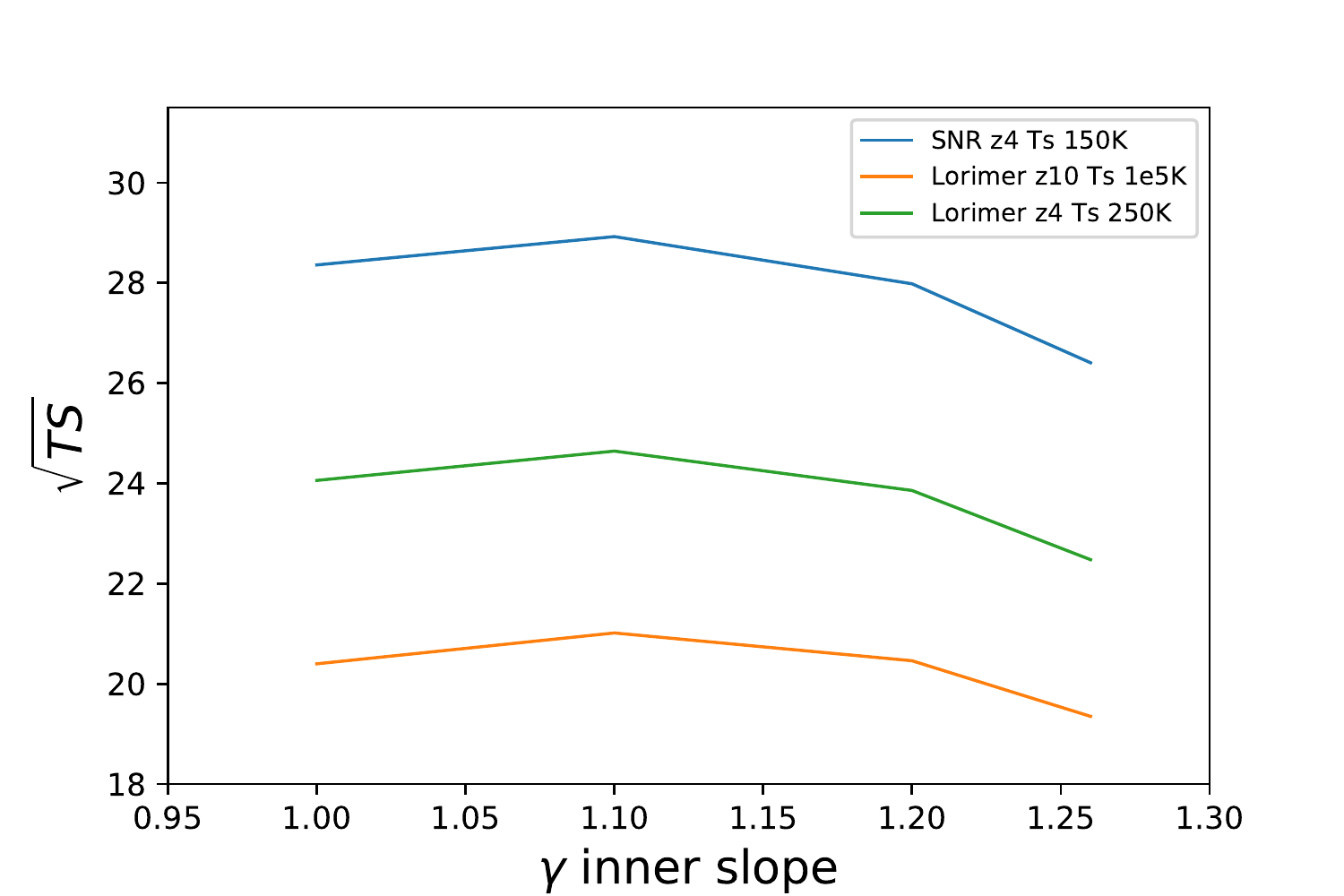}
  \caption{Best fit values for $\gamma$ using the three different IEMs for training the ConvNet. TS stands for Test Statistics and is defined as $TS=-2 \rm{ln}\Delta \mathcal{L} $, where $\Delta \mathcal{L}$ is the likelihood ratio between models with and without source, in our case the gNFW component \cite{2010ApJSTS}. The inclusion of a gNFW component always improves the quality of the fit. In this proof-of-concept work $\gamma=1.1$ as this value provides the larger Test Statistics (TS) over the different models tested. It is worth noting that TS  is not well defined for extended sources, we use this quantity as an approximated proxy for comparing models, but not to choose the best model, just to select one of them.}
  \label{fig:Delta_llike_Train_ABC}
\end{figure}

Standard {\it Fermi} tools are used to prepare data and to perform the fits. In this analysis we fit the following set of parameters: 
\begin{itemize}
\item The normalization of the inner rings in the IEM.
\item The normalization of the brightest 3FGL sources (we neither vary the position nor the spectral shape of the point sources, as this would increase the freedom significantly in the fitting). It is worth mentioning that we use a dataset derived with improved reconstruction algorithms and with more years of exposure than in the derivation of the 3FGL catalog.
\item We model the GCE spatially with a gNFW, spectrally as a WIMP of 50 GeV annihilating into $b\overline{b}$ quarks. In this way the only free parameter for this component is the normalization, that we parametrize as the thermally averaged cross section $\langle \sigma v \rangle$.
\end{itemize}
 The internal slope of the gNFW template is varied, changing the $\gamma$ parameter in eq~\ref{eq:gNFW}. For each value of $\gamma$, we run five different fits corresponding to the five IEMs selected. Inspired by the results of \cite{1704.03910} the following values for $\gamma$ are used: 1.0, 1.1, 1.2, 1.26, and 1.3. In the simulations used for training and validating the ConvNets, $\gamma$ is set at 1.1 as we find a slightly better fit with this configuration, see figure \ref{fig:Delta_llike_Train_ABC}. Table \ref{tab:diffuse} presents the likelihood of the five selected IEMs together with the resulting  $\langle \sigma v \rangle$ when the IEMs to generate the training and validation data uses gNFW with $\gamma=1.1$. It is worth noting that all IEMs were previously fitted to all-sky data and based on different assumptions about the initial parameter. Therefore the likelihood ratio test can not be used to compare models as they are not special cases of a null model. However, after inspection of spatial and spectral residuals, we get enough agreement to the data for this methodology analysis.
 
\begin{table}[t!]
\caption{Results of fitting the $15^o\times 15^o$ region around the GC using five different IEMs. The gNFW template is created with  $\gamma=1.1$. All likelihoods and both spatial and spectral residuals are at the same level, implying that all models provide a similar description of the data. We assume the sources of CRs are supernova remnants (SNRs) considering two CR
source distributions, one traced by the observed distribution of pulsars, Lorimer \cite{Lorimer:2006qs}, and other tracing SNRs observed \cite{Case:1998qg}. $T_S$ stands for spin temperature of the atomic hydrogen for the derivation of gas column densities from the 21-cm line data.}
  \begin{center}
   \begin{tabulary}{0.7\textwidth}{C}
    \begin{tabular}{| c | c | p{2cm} | l | c | c |}
    \hline
     Usage  &   CR distribution  &  Halo height $z$ (kpc)  & $T_S$ (K)   &  $\text{Log}\mathcal{L}$    & $\langle \sigma v \rangle \times 10^{-27}$ cm$^3$/s \\ \hline
     Training A &  SNR  &  10  & 150 & -442855 & 45.59  \\ \hline
     Training B &  Lorimer  &  10  & $1\times 10^{5}$ & -442304 &  33.61  \\ \hline
     Training C &  Lorimer  &  4  & 150 & -442357 & 39.32  \\ \hline
     Testing A &  Lorimer  &  10  & 150 &  -442539 & 39.63  \\ \hline
     Testing B &  SNR  &  4  & $1\times 10^{5}$ & -442664 & 42.67  \\ 
     \hline
     \end{tabular}
    \end{tabulary}
    \end{center}
\label{tab:diffuse}
\end{table}

\begin{figure}[t!]
\begin{center}
\begin{tabular}{ll}
 \hspace*{-10mm}\epsfig{file=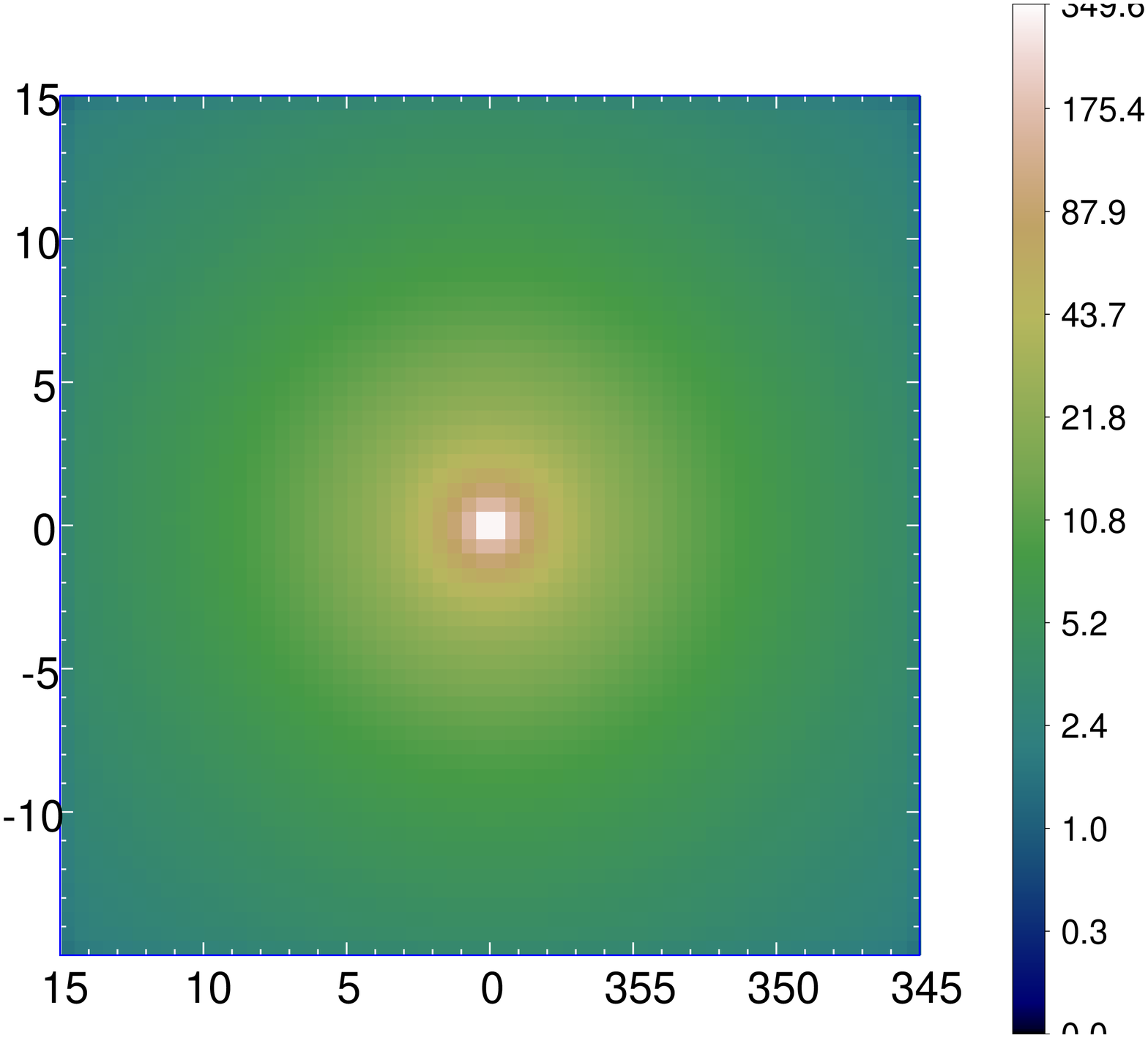,height=5.0cm} & \epsfig{file=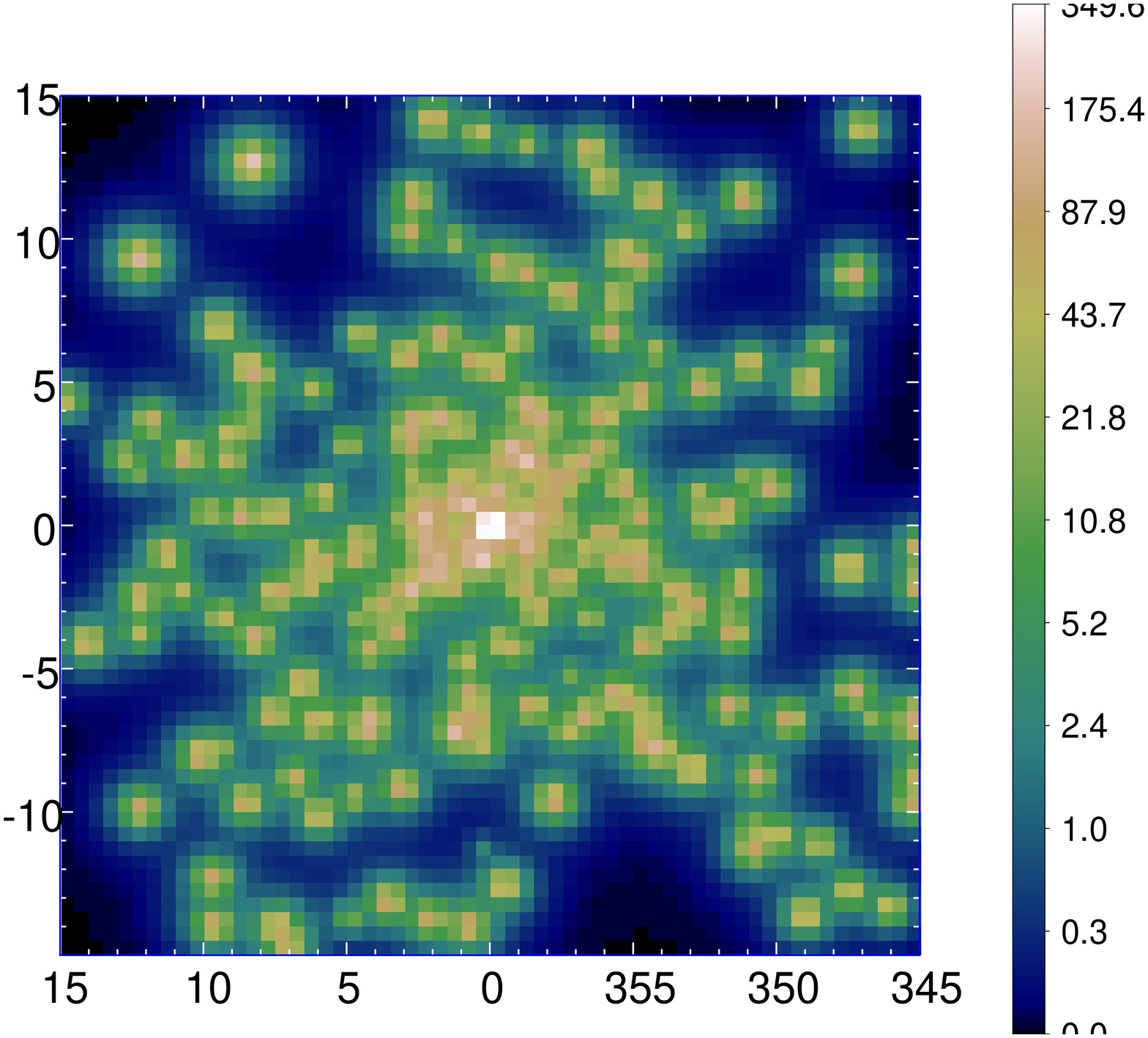,height=5.0cm} \\
\end{tabular}
\end{center}
\caption{Count maps in the 1-6 GeV energy range comparing the granular vs diffuse nature of the GCE, these templates are indistinguishably for the template method (see section \ref{sec:fit_procedure_results}). Both maps have the same total emission and follow a gNFW distribution.}
    \label{fig:comparison}
\end{figure}

\subsection{Simulation of { \it Fermi }-LAT data}
\label{sec:simulating-fermi-data}

After fitting the five models of the Galactic Center region described above to the {\it Fermi}-LAT observations, we generate simulated data using the gtsrcmap and gtmodel codes of the {\it Fermi} tools, the former for the convolution of models with the {\it Fermi}-LAT response, the later to combine diffuse templates and generate the point source populations. We add Poisson noise to the final images. In this set of fabricated images, we modify the fraction $f_{\rm src}$ between cumulative emission from point sources and a diffuse component of the GCE model, keeping its magnitude, spectral shape and distribution as determined by the fits, see figure~\ref{fig:comparison}.

\begin{figure}[t!]
\begin{center}
\epsfig{file=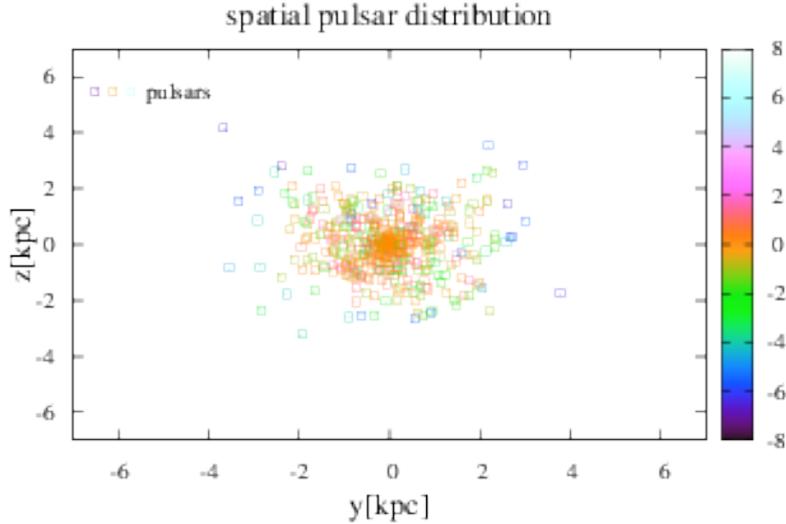,height=8.5cm} 
\end{center}
\caption{3D distribution of a point source distribution following an gNFW profile with $\gamma=1.1$. Each time such a population is generated the sources are placed at different positions. The flux $s$ of each source is also randomly assigned. The (0,0,0) point is the GC, the color code stands for $x$ axis in kpc.}
    \label{fig:population}
\end{figure}

The spectral shape of the GCE is highly dependent on the assumptions of the underlying model, but as can be noted in Figure \ref{fig:GCE} the excess is always present in the 1-6 GeV band. We, therefore, choose that band to generate images for training and testing the ConvNets. The left panel of figure~\ref{fig:comparison} shows an image made with the gNFW template ($\gamma=1.1$) in the 1-6 GeV band. The right-hand side of figure \ref{fig:comparison} shows the image of a simulated population of point sources that produce the same amount of total flux as the gNFW template. The density of sources in the population is also spatially distributed following equation~\ref{eq:gNFW} with $\gamma=1.1$. To generate the maps of point source populations which mimic the GCE, we make the following assumptions: all sources have the same spectral shape and are distributed randomly following a density distribution specified above (see figure~\ref{fig:population} for a sample plot of a randomized population of point sources). Besides following equation~\ref{eq:gNFW}, we assume a power-law behavior for the flux distribution in the population,

\begin{equation}\label{flux}
\frac{dN_{src}}{ds} = A s^{\alpha}.
\end{equation}
The normalization $A$ is determined by $f_{\rm src}$ times the total emission of the GCE as obtained in the corresponding fit\footnote{$\int_{s}{As^{\alpha}ds}$ equals to $f_{\rm src}$ times total GCE flux. The GCE flux varies with the background used in the fit, therefore depending on the background and fraction $f_{\rm src}$ we determine the parameter $A$.} and the constraint that all sources are below the 3FGL detection threshold ($1\times 10^{-9}$ cm$^{-2}$ s$^{-1}$ for fluxes above 1 GeV \cite{Acero:2015hja}). $s$ represents the source flux and $\frac{dN_{src}}{ds}$ the number of sources for a specific flux interval. The spectral index $\alpha$ determines the number of point sources per flux below the 3FGL detection threshold. Thus, a high value of $\alpha$ means most point sources are just below to the 3FGL detection threshold, while a low value of $\alpha$ means most point sources are far below the 3FGL detection threshold. For example, a value of $\alpha=1.05$ implies that almost all sources have a flux close to the detection threshold. In this analysis we let $\alpha$ range from  -1.05 to 1.05\footnote{In \cite{Fermi-LAT:2017yoi} the best-fit $\alpha$ value found for the Bulge pulsar population is -1.2. We do not extend our $\alpha$ range to cover that value as we only work with simulations, follow-up work will be devoted to extracting physical quantities from actual {\it Fermi}-LAT images.}.

\begin{figure}[t!]
\begin{center}
\epsfig{file=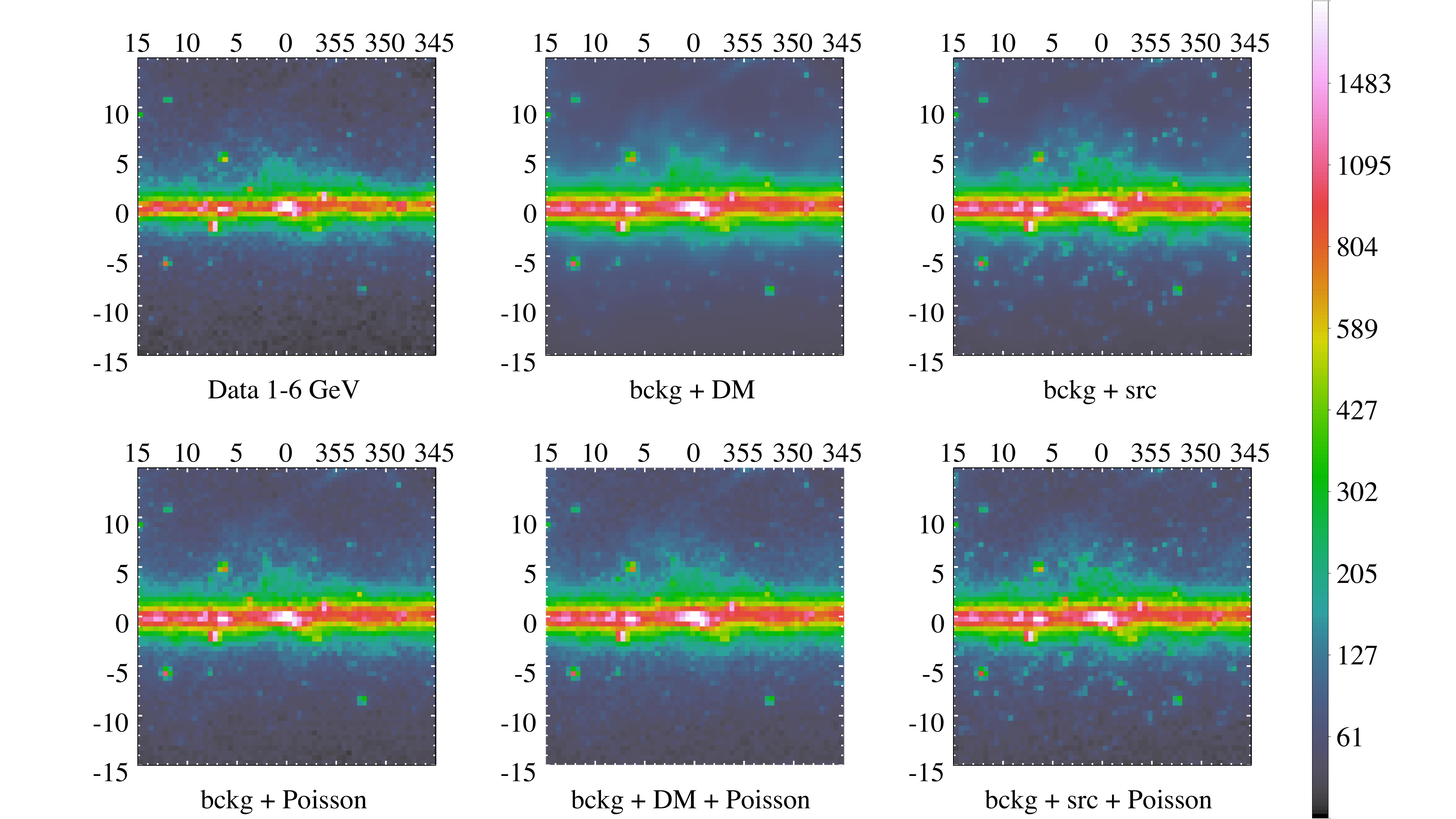,height=8.5cm} \end{center}
\caption{Images of the Galactic center in the 1-6 GeV $\gamma$ ray band. Color code in counts/pixel. Upper left: real data. Upper center: a model of the GC region as result of the fit to the Training A model. In the upper right the same model but a distribution of point sources replace the gNFW template. The lower row presents the same models of the upper row but with Poissonian noise added to make them more realistic. We use the gtmodel code of { \it Fermi } Science Tools version 11-05-00 to create the images. \it{Bckg} stands for background, which is emission except for the GCE. \it{DM} for the diffuse component and \it{src} the point source population below the detection threshold.}
    \label{fig:results_training}
\end{figure}

In figure~\ref{fig:results_training} we present some of the models for training, focusing on the extreme cases where the GCE is entirely made by DM annihilation or by a point source population. 
We generate training and testing data as follows: first, we convolve the best-fit model with the instrumental response function of the {\it Fermi} LAT (using the gtmodel tool of the Fermi science tools\footnote{\url{https://fermi.gsfc.nasa.gov/ssc/data/analysis/scitools/overview.html}}) yielding sky images with the expected number of counts. Then we generate Poisson instances of these pictures.

One of the main concerns of this approach is the quality of the simulations versus the actual {\it Fermi}-LAT image of the Galactic center. If the simulations are not representative of the real data, the output is unreliable even if the ConvNet's accuracy is 100\%. The 'Reality Gap' is the name of this discrepancy between simulated and real data.  We use for simulations state-of-the-art modeling of the diffuse emission to represents the actual data. However, some model parameters that serve as input for the simulation are not well determined by our best-fit procedure. These include:

\begin{itemize}
\item Which IEM is correct? The computation of IEMs requires input data that are highly uncertain, the models of our Galaxy represent quite well the whole $\gamma$ ray sky, but in the inner region, the complexity of the environment is significantly higher \cite{Ackermann:2012pya}. In particular, GALPROP-generated templates are 'smoothed' versions of reality as many smaller molecular clouds are not present in the GALPROP-input gas maps. For example, the small-scale structure of the IEM is not present in the training set, see Supplemental Material, item F, of \cite{2016PhRvL.116e1102B}. Furthermore, the models in reference~\cite{dePalma:2015tfa} do not enclose the complete range of systematics involved in the modeling of Galactic diffuse emission. They only consider one method for creating the IEMs and varied three input parameters. For a ConvNet to give accurate results on the real data, the variations of the input parameters need to be large enough so that it can learn to generalize over those parameters.
\item The list of detected point sources: we used the 3FGL catalog, which was created using four years of data, while we use about seven years of data to generate the simulations. As new sources are expected to appear with more data, the results of the network are going to be biased towards higher $f_{\rm src}$. Therefore, sources that could be detected with seven-year data, but not present in the four-year catalog, will in the simulations be part of the signal while they should be included in the background. This miss modeling leads to a higher predicted $f_{\rm src}$ value by the network. The 2FIG catalog presents about 200 more sources detected in the GC region than the 3FGL catalog and used seven years of data, which will remove this bias. For this initial analysis we only wanted to create images similar to the GC but in a simple way, newer catalogs will be part of the models in a follow-up study.
\item The positions of point sources part of the unresolved population: we fix only their distribution in the current simulation.
\item The true distribution of the GCE emission: in our work we use gNFW template parametrized by $\gamma$ in equation \ref{eq:gNFW}, which is here fixed to 1.1.
\end{itemize} 

We take the following action to account for these unknowns:\\

The network was trained using simulations of three background models (training A, B, and C in table \ref{tab:diffuse}) while validating the ConvNets on models that were generated using two different background models (testing A and B in table \ref{tab:diffuse}). In total five background models were used. This procedure ensures that the network is sensitive to features common to all background models instead of training on characteristics of one particular background model only.\\

We use many initializations of the sources, and in every model simulation, their actual positions were different. Therefore the network cannot rely on fixed positions of the point sources, but only on their distribution which is determined by the parameter $\alpha$ (see section \ref{sec:analysis_setup}).

\section{Neural network setup}\label{sec:ConvNetsetup}

This section describes the analysis setup of the convolutional neural networks that were used to make the predictions of $f_{src}$. The input data of the network is a $(w,h,c)$-tensor, with $w$ and $h$ the width and height of the image and $c$ the number of channels (colors in case of a color image). As the pixels of the image represent photon counts in the band between 1 and 6 GeV, the number of channels is one (meaning the image is monochromatic). The output of the network is a number between 0 and 1, representing the value of $f_{\rm src}$. In total five networks were trained with the same network architecture, see table \ref{tab:network}. We train one network on the full image, and other four networks trained on half of the image (left, right, top and bottom half). After training, the output values of the networks trained on half of the images can be compared to check if their predictions are in agreement. Also, the output values of these four networks can be averaged to obtain an averaged prediction of these four networks. Averaging multiple network results typically improves overall accuracy \cite{ensamble} (also called \textit{ensemble learning}).

\begin{table}[h!]
\centering
\caption{The five different networks trained on the simulations.}
\label{tab:network}
\begin{tabular}{ll}
\textbf{Network name} & \textbf{Tensor size}     \\
Full image            & $(120,120,1)$            \\
Left half of image            & $(60,120,1)$             \\
Right half of image           & $(60,120,1)$             \\
Top half of image             & $(120,60,1)$             \\
Bottom half of image          & $(120,60,1)$             
\end{tabular}
\end{table}

We average the predictions of the ConvNets trained on partial images; the result is labeled 'averaged network' in what follows. The training data contains images that are generated using three different background models. The validation data consists of 60.000 images from two background models that are not part of the training data (Training A and B in table \ref{tab:diffuse}, we split the GC pictures evenly) and with randomized $\alpha$ and point source locations. Therefore, to get high accuracy with the ConvNet recovering $f_{\rm src}$ on the 60.000 images in the validation set, we must account for variations in the locations and fluxes of the individual unresolved sources, as well as uncertainties in the IEMs.

As a preprocessing step, we normalize the pixel values of the images to values between 0 and 1 to improve the training speed. The last layer of the network architecture (see Appendix \ref{appendix:convnets}) has a sigmoid activation function, $\sigma(x)=(1+e^{-x})^{-1}$. This function goes asymptotically from 0 to 1, making an $f_{\rm src}=1$ prediction impossible (the input of the last layer has to be infinite in this case). To improve network accuracy for very high $f_{\rm src}$ predictions, the output value of a prediction is normalized with a factor of $\frac{1}{\textrm{max}(f_{\rm src})}$ to map the highest predicted output to one. Here $\rm max(f_{\rm src})$ is the highest predicted value of the validation data.

\subsection{Network architecture}
\begin{figure}[h!]
\begin{subfigure}{\textwidth}
  \centering
  \includegraphics[width=\textwidth]{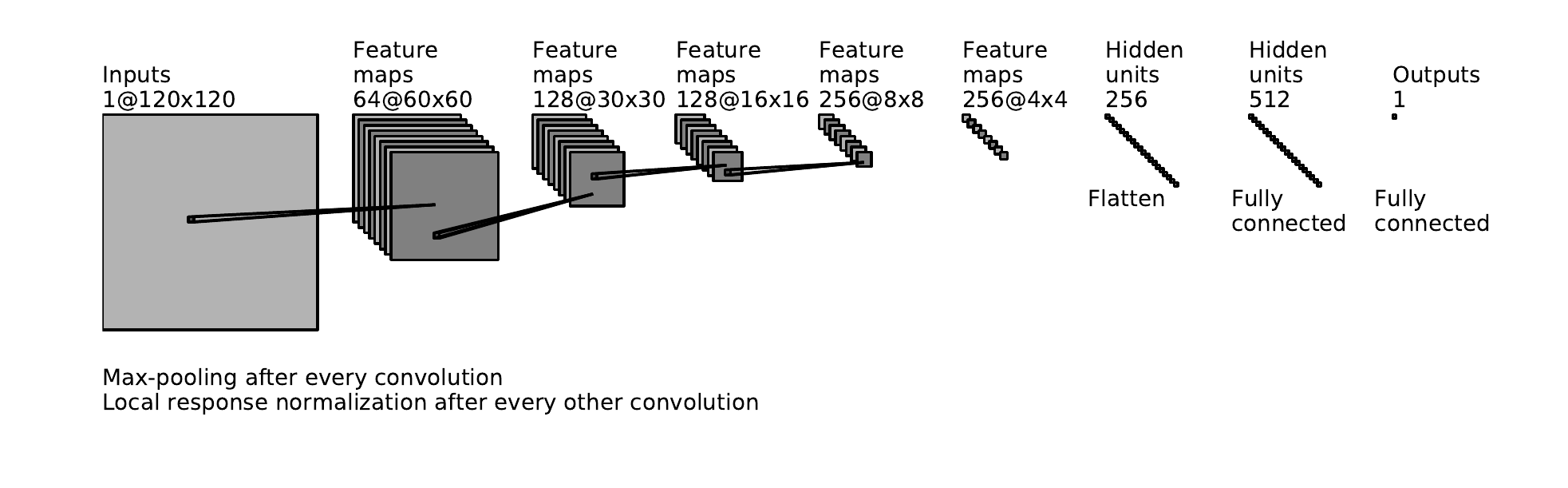}
\end{subfigure}
\caption{Visualization of the convolutional neural network. The network consists of an input layer, 5 convolutional + pooling layers, 2 fully connected layers and finally an output layer. See Appendix \ref{appendix:convnets} for an introduction on ConvNets.}
\label{fig:convnet}
\end{figure}

This section explains the architecture of the ConvNet used for this work. For a general high-level introduction to ConvNets, see Appendix \ref{appendix:convnets}. The ConvNet architecture is identical for all five networks and is visualized in figure \ref{fig:convnet}. The activation functions of all inner layers are ReLU functions \cite{icml2010_NairH10}: $f(x)=x$ if $x>0$ and $f(x)=0$ otherwise. The last layer has a sigmoid activation function.

The loss function of the network is mean square error; the optimizer is chosen to be the Adam optimizer \cite{DBLP:journals/corr/KingmaB14}. We train the ConvNet in two steps: first with a learning rate of $10^{-5}$ and then with a learning rate of $10^{-9}$, both using 20 epochs \cite{DBLP:journals/corr/KingmaB14}. We used the TensorFlow\footnote{https://www.tensorflow.org} library and ran training on two Nvidia GTX1080 cards.
\section{Results}
\label{sec:results}





\subsection{Neural network results}

\begin{figure}[h!]
\centering
\begin{subfigure}{\textwidth}
  \centering
  \includegraphics[width=\textwidth]{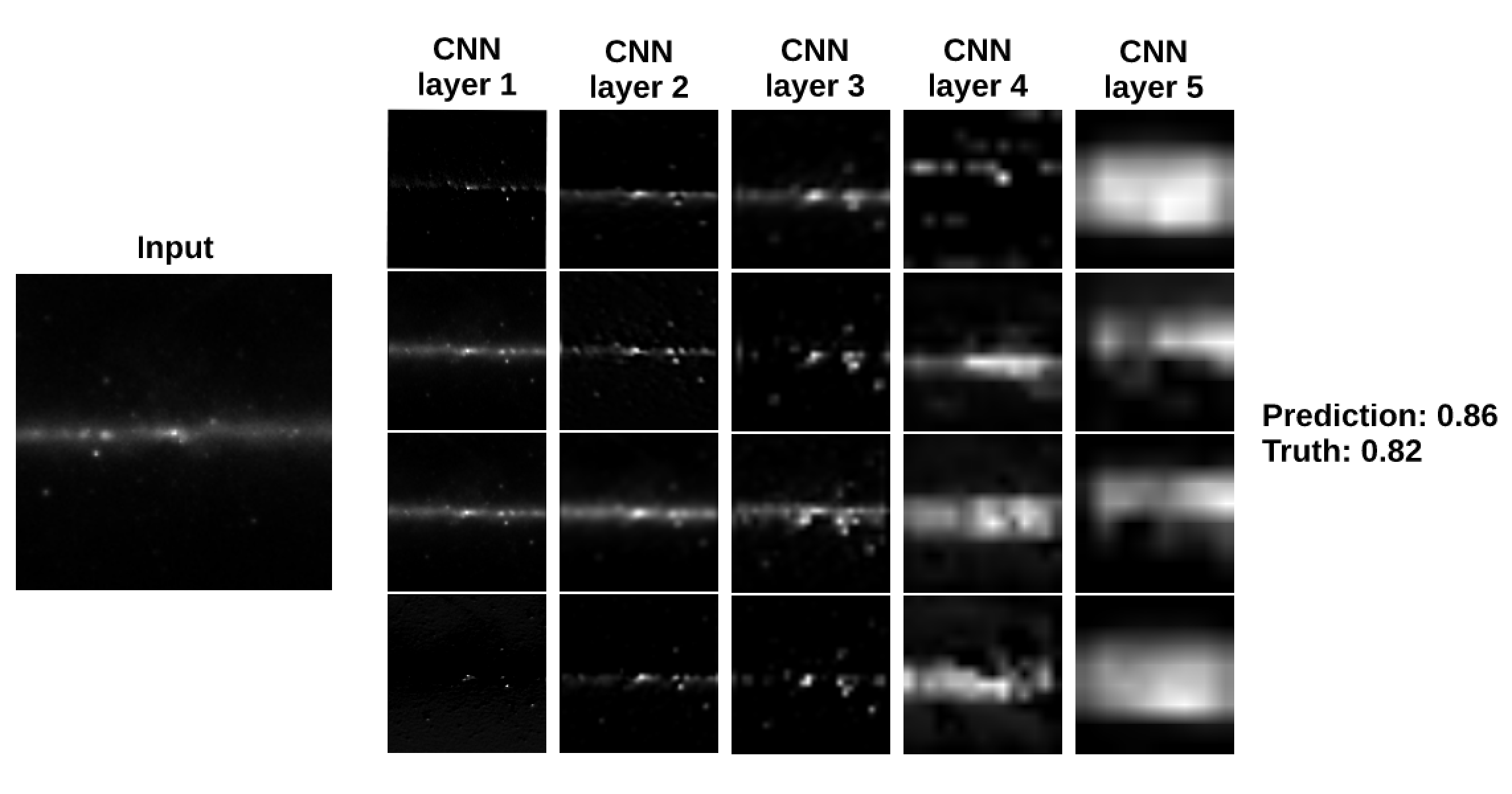}
\end{subfigure}
\caption{Example filters of different layers of the full network on a simulation. Each column of images represents a convolutional layer. The data flows from left to right (from input to prediction). Each convolutional + pooling layer accepts the image as input and outputs $n$ smaller images, where $n$ is the number of convolutional filters in that layer ($n$ ranges from 64 to 256, see figure \ref{fig:convnet}). Only four filters per layers are shown for clarity.}
\label{fig:conv-output}
\end{figure}

Figure \ref{fig:conv-output} shows a visualization of the different activations within the ConvNet layers. The input image on the left. Each column represents a convolutional layer, and the image passes through the network from left to right. Only four convolutional filters are shown per column for clarity. In the actual network, the number of filters varies between 64 and 256. Also, because there is a max pooling layer (see Appendix \ref{pooling}) between the convolutional layers, the images are smaller in subsequent convolutional layers. We zoomed them to identical sizes in the visualization. The figure shows that some filters tend to filter out the diffuse background, while others seem to take the diffuse background into account only. Internally the network seems to decompose images into its diffuse and point source components. For this particular simulated sample, the difference between the ConvNet prediction and the actual value of $f_{\rm src}$ of the sample was 0.04.

\begin{figure}[h!]
\centering
\begin{subfigure}{.5\textwidth}
  \centering
  \includegraphics[width=\textwidth]{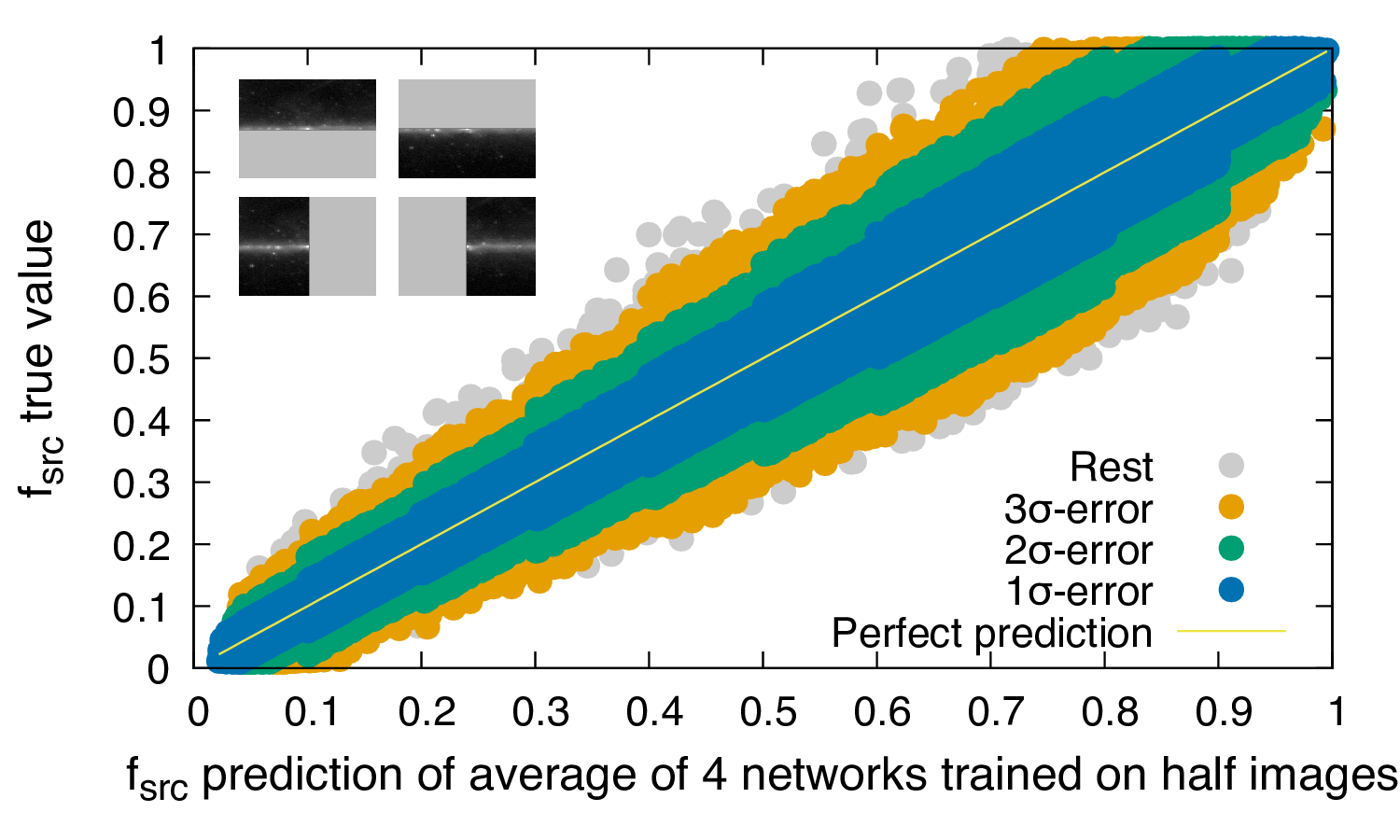}
  \caption{Prediction of the average network\\versus true values.}
  \label{fig:allAvg_truth}
\end{subfigure}%
\begin{subfigure}{.5\textwidth}
  \centering
  \includegraphics[width=\textwidth]{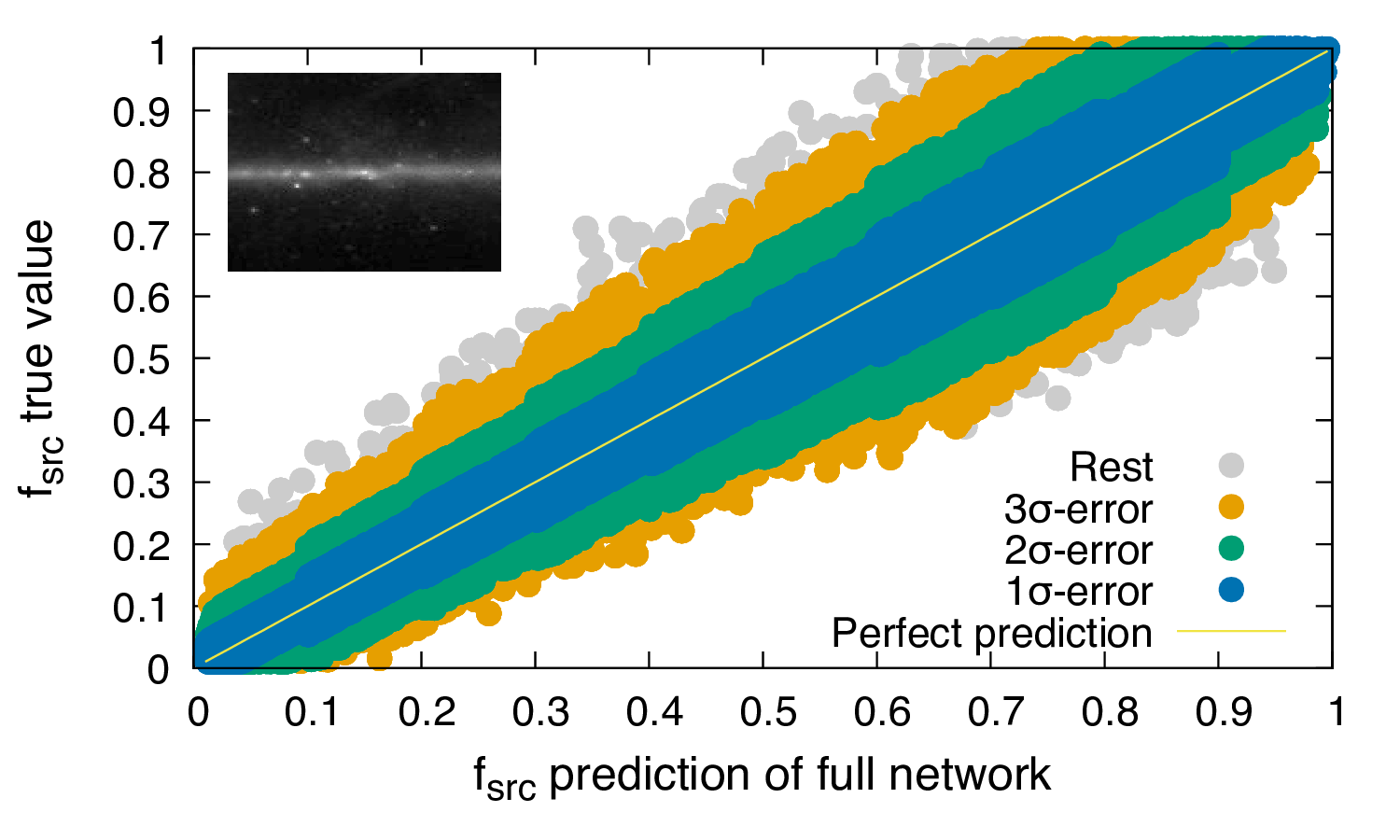}
  \caption{Prediction of the full network\\versus true values.}
  \label{fig:full_truth}
\end{subfigure}
\begin{subfigure}{.5\textwidth}
  \centering
  \includegraphics[width=\textwidth]{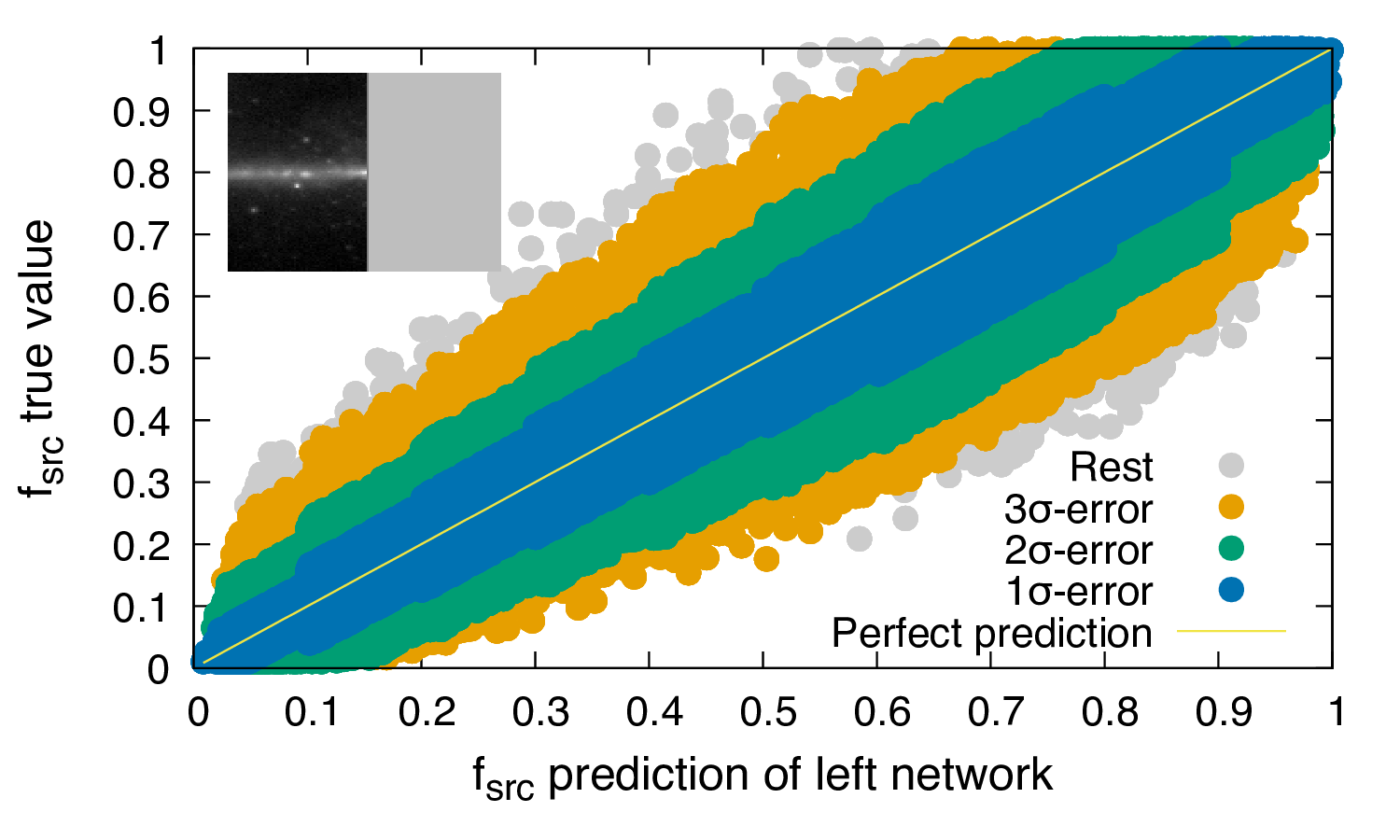}
  \caption{Prediction of the network trained on the left half versus true values.}
  \label{fig:left_truth}
\end{subfigure}
\caption{Network results on the validation set. The different colors represent the $1\sigma$, $2\sigma$ and $3\sigma$ bands. The diagonal line represents a perfect prediction.}
\label{fig:results}
\end{figure}

In figure \ref{fig:results} we show the results of the ConvNet output on the validation data for three different networks: the full network, the averaged network and the network trained on the left half of the image. Note that these are predictions on two different models that were not in the training data (Testing A and B models in table \ref{tab:diffuse}). We compute the accuracy of the network in 10 equally spaced bins based on this validation set. This is done because the network is more accurate for low $f_{\rm src}$-predictions than for high $f_{\rm src}$-predictions (see figure \ref{fig:results}). The average network performs slightly better than the full network, meaning an ensemble of different networks achieves higher accuracy than training one network on the full image (see section \ref{sec:ConvNetsetup}).\\

\begin{table}[h!]
\centering
\caption{Properties of the network trained on the left half at the predicted value on the real { \it Fermi } result.}
\label{tab:real-output}
\begin{tabular}{ll}
\textbf{Name}    & \textbf{Value} \\
Predicted output & 0.887          \\
$1\sigma$ error  & 0.105          \\
$2\sigma$ error  & 0.210          \\
$3\sigma$ error  & 0.324          \\
Maximum error    & 0.416         
\end{tabular}
\end{table}

To minimize the researcher bias for the future implementation of this method only the network trained on the left half of the image was used to predict the real {\it Fermi}-LAT data. The output of this ConvNet is $0.887$, putting the network value in the bin with the errors shown in table \ref{tab:real-output}. We reserve the other half and the full picture for the follow-up work.\\

The predicted value for $f_{\rm src}$ is $0.887 \pm 0.105$. This value favors an interpretation regarding point sources for the GCE. However, it is worth noticing that we must not use this ConvNet on real data as the images used for training and testing the ConvNets use the 4-years 3FGL catalog while we use about seven years of data. The newer Second Fermi Inner Galaxy catalog (2FIG) contains about double the sources in the same region \cite{Fermi-LAT:2017yoi}. Also, we train the networks on $\alpha$ values between -1.05 and 1.05, but larger values are physically plausible. This will be addressed in a follow-up paper.

\begin{figure}[h!]
  \centering
  \includegraphics[width=\textwidth]{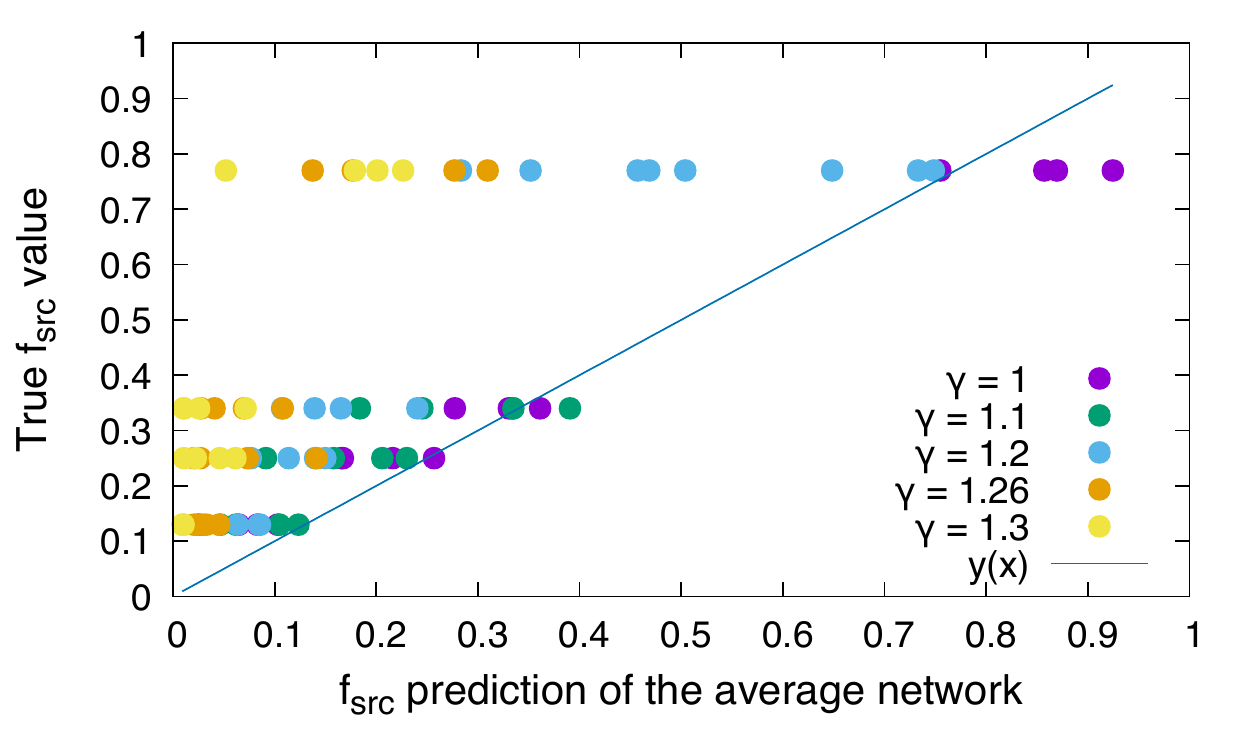}
  \caption{Predictions of the average network using different values for $\gamma$. For $\gamma=1$ and $\gamma=1.1$ the prediction are still accurate. However for higher values of $\gamma$ the network starts to under-predict the true values of $f_{\rm src}$. This shows that the network does not generalize over higher than $\gamma=1.1$ values, but does for values between 1 and 1.1.}
  \label{fig:gamma_test}
\end{figure}

We use a fixed value of $\gamma$ in the simulations. To see if changing $\gamma$ negatively impacts the predictions few simulations were run with varying $\gamma$ values. As can be seen in figure \ref{fig:gamma_test} this had a sizable impact on the result. For  $\gamma=1.1$ and $\gamma=1$ the network predictions remain accurate, but for higher values of $\gamma$ the network under-predicts $f_{\rm src}$. Because a higher value of $\gamma$ means that the point sources are closer to the GC, it becomes increasingly difficult to distinguish the emission from the point-source population from truly diffuse emission. It is therefore expected that the network under-predicts $f_{\rm src}$ for higher values of $\gamma$. Since the most likely value of $\gamma$ we obtained in our fit is between 1 and 1.1, it is expected that the network is still accurate on the real {\it Fermi}-LAT data. However, the next iteration of the network will be trained on multiple values of $\gamma$ to ensure that the $f_{\rm src}$ predictions are accurate regardless of different $\gamma$ values.

\section{Conclusion}
\label{sec:conclusion}

Training a convolutional neural network on simulated data of the {\it Fermi} LAT yields promising results to characterize the contribution of unresolved dim point sources to the diffuse emission. We find that the ConvNet method gives predictions beyond the classification of either 100\% point sources or diffuse source: it can also predict a mixture. The error on the prediction of $f_{\rm src}$  is usually on the order of 10\% for the averaged network. Furthermore, this method is not limited to predicting $f_{\rm src}$. One can also make predictions on the properties of the source population, like $\alpha$ and  $\gamma$, as well as the most likely background model by training networks on those particular values.

When applied to the real data, the left network prediction of the point source fraction $f_{\rm src}$ of the Galactic center excess is $0.887\pm0.105$. This value disfavors the possibility that the GCE is solely caused by diffuse emission but rather by a mixture of combined radiation of unresolved point-sources and genuinely diffuse emission, in agreement with previous studies~\cite{Fermi-LAT:2017yoi,2016PhRvL.116e1103L,2016PhRvL.116e1102B}. The present study only presents a proof of concept, and one should note that the results are biased towards higher values of $f_{\rm src}$. The reason for this bias is the use of a four-years catalog of point sources, the 3FGL, whereas in this analysis we use seven years of data. More recent lists of point sources, such as the 2FIG, already contain twice as many objects in the same region. Also, to extend the range of the $\alpha$ parameter from $\pm 1.05$ to at least $\pm 1.7$ to account for best-fit values found by other studies~\cite{Fermi-LAT:2017yoi,Bartels:2017xba}.

Currently, the training data of the network is a value per pixel which represents the number of photon events between 1 and 6 GeV. However, we can feed more information into the network to make more accurate predictions and allow for more generalizations. These improvements include multiple values of $\gamma$, as discussed in the previous section. Instead of a $(w,h,1)$-tensor representing the counts of 1-6 GeV photons, the next iteration of this network will use a $(w,h,c)$-tensor, where $c$ is the number of energy bins. Each bin will contain the pixel count of some specific energy bin. With this extra dimension of the data, the network can learn between correlations in any direction of the tensor. This extra information may lead to improved performance and will allow more complex architectures. The full analysis will yield the following improvements:
\begin{itemize}
\item Increase sensitivity of $\alpha$ to realistic values from $\pm1.05$ to at least $\pm1.7$
\item Generalize over multiple values of $\gamma$ instead of assuming $\gamma=1.1$
\item Use the latest point source catalog instead of the 3FGL
\item Use multiple energy bins as channels instead of one channel
\end{itemize}

\begin{appendices}
\section{ConvNets}
\label{appendix:convnets}

Recently the field of deep learning has received a lot of attention because of the predictive power of deep neural networks. These networks can be used to detect objects in images~\cite{NIPS2012_4824}, create text-to-speech algorithms \cite{DBLP:journals/corr/OordDZSVGKSK16} and many more applications.  This novel approach utilizes the many advances in machine learning of the past years and can lead to better predictions using fewer assumptions, as ConvNets typically require the raw data as input. Using DNNs over conventional data analysis methods has up- and downsides:

\begin{itemize}
    \item DNNs work with raw data and learn to recognize correlations in the data automatically. This has two advantages: 1) there is no need to manually prune the data (manually denoise, reparametrize or otherwise reduce the dimensionality of the data) and 2) the raw data contains all possible information. Pruned data does not. A conventional method that uses preprocessed data cannot access all the information enclosed in the raw data. The quality of the pruned data is entirely dependent on the human understanding of the data and it might happen one unknowingly removes correlations when preprocessing data.
    \item DNNs are very general. The network architecture changes from problem to problem, but the layer types and methods can be used in many different problem areas like analyzing image or video data, regression, speech analysis and many more domains. This property of DNNs means the advances in one area of research (for example face detection) also benefit all the other research areas.
    \item DNNs can generalize over randomness. When a DNN needs to learn magnitudes of stars in a 2D image, the randomized location of the star carries no information. By using convolutional layers and enough training samples, the network can learn to 'ignore' the location of the star. This generalization is an important feature in the analysis in this work: the unresolved point sources have a random location in our simulations.
    \item DNNs are hard to train. The training process tunes many hyperparameters to specific values for the network to make accurate predictions. These hyperparameters need to be set before training and require knowledge of the problem, the network, and trial-and-error.
    \item DNNs require a lot of data and processing power. This is one of the main reasons DNNs are only becoming popular in recent years. The hardware required to train networks that can make accurate predictions on real-world data is only available since the advent of the GPU. Also, deep neural networks typically require a lot of training examples to train well.
\end{itemize}


\subsection{ConvNet data pipeline}

\begin{figure}[t!]
\begin{center}
    \epsfig{file=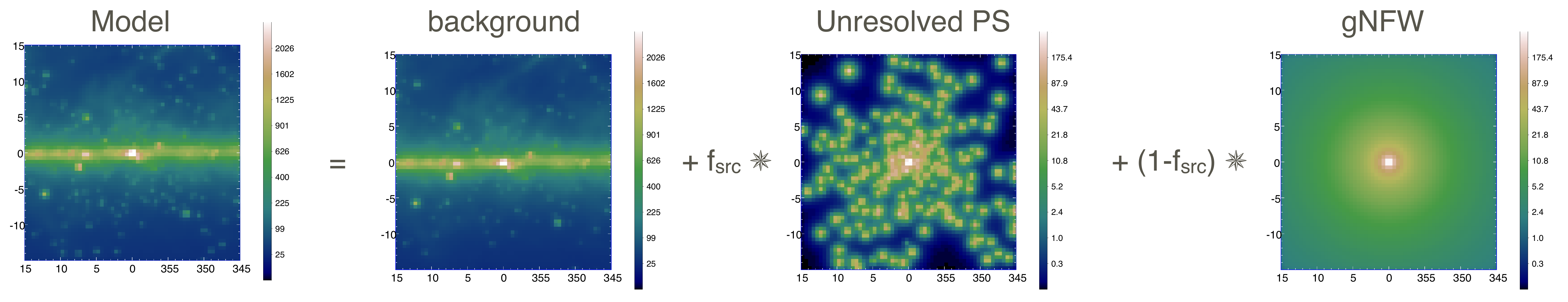,height=3.0cm} 
\end{center}
 \caption{Visualization of the model construction. The model (that is used as training data for the ConvNets) is generated by adding a fraction $f_{src}$ of point sources and a fraction of $1-f{src}$ of the diffuse source to the background. The sum of the diffuse and point source fraction that is added on top on the background always equals the GC excess.}
  \label{fig:cartoon}
\end{figure}

In general, a ConvNet takes an N-dimensional input, transforms it using different layers, and produces an M-dimensional output. This can be a prediction that a specific input belongs to a particular class (e.g., object detection) or a prediction of a regression problem (e.g., this research). The input of a computer vision problem is typically an image. The image can be represented by a $(w, h, c)$-tensor, where $w$ and $h$ represent the width and height of the network and $c$ the number of channels. For instance: a $120\times 120$ color image has $120 \cdot 120 \cdot 3 = 43.200$ different values embedded into it. In this analysis, the number of channels is one, and it represents the photon count in the energy bin 1-6 GeV. The input is the first layer of a neural network and in the example consists of $43.200$ neurons. When an image is fed into the network, each input neuron consists of one color of one-pixel value. In left-hand side of figure \ref{fig:cartoon} we present an example of an image for training. It is composed of a background emission, and the GCE, whose granularity is controlled by the {\it fraction of point sources } $f_{\rm src}$ parameter, which varies between 0 and 1. A value of $f_{\rm src}=0$ means the GCE is composed of only a diffuse source, and $f_{\rm src}=1$ means the GCE is composed of only point sources. With ConvNets we transform such images into predictions of the parameter $f_{\rm src}$. 

In ConvNets, between the input and output layers, there are so-called {\it hidden layers}. In this work we use the following kinds of hidden layers: fully connected, convolutional, max pooling and local response normalization. The differences of these layers will be discussed in the next section.

\subsubsection{Layer Architectures}\label{pooling}

\begin{figure}[h!]
\begin{center}
    \includegraphics[width=0.4\textwidth]{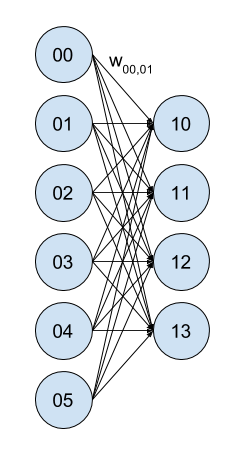} 
\end{center}
 \caption{Schematic view of a fully connected layer with 6 input and 4 output neurons. The 24 lines represent the weights of the connections between the the neurons from the two layers.}
  \label{fig:fclayer}
\end{figure}

A fully connected layer connects all neurons of a layer to the next one. A weight $w$ is assigned to each connection (see figure \ref{fig:fclayer}). The value $n_{y,j}$ of a particular neuron $j$ in layer $y$, is defined as $n_{y,j}=f\left( \sum_i n_{x,i} \cdot w_{x,i;y,j} + \textrm{bias} \right)$, where $f(x)$ is called the activation function. In this research the internal layers have a Rectified Linear Unit (ReLU) activation function ($f(x>0) = x$ and $f(x\leq0) = 0$). The output layer has a sigmoid activation function, to map all input values to the range (0,1), which represents the output range of  $f_{\rm src}$ parameter.

During training, an image is fed into the network, and an $f_{\rm src}$ output value is predicted. This output can be compared to the actual $f_{\rm src}$ used to create the image. The error is propagated back into the network which {\it adjusts} all $w$ and biases \cite{LeCun:1998:EB:645754.668382}. Before training all weights and biases are set to normal distributed random numbers (the network initialization).

A DNN with only fully connected layers has many parameters (weights and biases) to train. For instance: a 120x120 color image has 43.200 input neurons. If the neural network has one hidden layer with 100 neurons and one output neuron, the number of weights in the ConvNet is $43.200 \cdot 100 + 100 \cdot 1 = 4.320.100$. This makes fully connected networks very hard to train.

A convolutional layer is a layer that has fewer weights than a fully connected layer by utilizing the fact that there is a correlation between neighboring pixels on an image \cite{lecun_bottou_bengio_haffner_1998}. Max pooling layers take a patch of an image and return the maximum of this value \cite{Zeiler2014}. A max pooling layer is used to reduce the dimensions of an image during network propagation. After multiple convolutional and pooling layers, fully connected layers are added to make a final prediction of the image. A graphical representation of a typical convolutional neural network can be found in figure 2 of \cite{lecun_bottou_bengio_haffner_1998}.

\subsubsection{Choice of activation function}

The choice of the activation functions of the neurons is crucial, as this introduces a non-linearity in the network and the right activation function optimizes the speed and predictive power of the network. If the activation function would be linear, for example, $f(x)=x$, the neural network would just be a linear combination of linear functions. If this is the case, it is impossible to make predictions about non-linear problems. Popular choices are the sigmoid and hyperbolic tangent (tanh) function because they are bound between 0 and +1 for the sigmoid function and -1 and +1 for the tanh function. However, they suffer from the so-called vanishing gradient problem \cite{Hochreiter01gradientflow}: the gradient at significant positive or negative values go to zero. As the gradient of the activation function is a used in the backpropagation algorithm to update the weights, the network effectively stops learning when the gradient is (almost) zero.

The ReLU activation function is currently a popular choice, because it has no vanishing gradient and is a non-linear function\cite{icml2010_NairH10}. On top of that, the derivative of the ReLU function is computationally trivial, while calculating the derivative of a sigmoid or tanh function can be computationally expensive. However, the ReLU function has no gradient for negative values, leading to so-called 'dying ReLUs': if the update of a weight makes the output of the ReLU negative, the particular neuron is effectively dead as there is no gradient anymore. This means the weights will not be updated anymore during training. There are many more activation functions that can be chosen and each of them has advantages and disadvantages. In this work, the ReLU activation function was chosen because of its speed and non-vanishing gradient.

\subsection{Training a neural network}
\label{appendix:training}

A neural network has many weights in it, which are initialized randomly according to some distribution (typically normal distributed). These weights need to be changed from the random initialization to particular values, to predict the training dataset accurately. This is done during the training phase of the network. During training, the network receives an image and calculates an (incorrect) output. The error is propagated back into the network, and the weights are set so the error of the prediction is less when the image would go through the network again. There are many backpropagation algorithms, of which the Adam optimizer is a popular one and the one used in this research \cite{DBLP:journals/corr/KingmaB14}.

A typical difficulty of training neural networks is overfitting. In this case, the network can predict every example in the training data perfectly, while it fails to generalize over the important features. Instead of learning those, it 'memorized' the training data and cannot be used for other cases. To determine whether a network is overfitted to the training data, one can set aside a part of the dataset and not use it during training. After the network is trained, the accuracy of the network on this validation set is calculated. It is an excellent indication that the network has overfitted if the training set accuracy is much higher than the validation set accuracy \cite{NIPS2000_1895}. There are a number of methods to prevent overfitting, such as regularization \cite{Ng:2004:FSL:1015330.1015435} and dropout layers \cite{Srivastava:2014:DSW:2627435.2670313}. In this research, the convolutional layers have an L2 regularizer to penalize large weights. Adding dropout has been considered as well, but only using L2 regularizers was enough to prevent overfitting and adding a dropout layer did not improve the network accuracy.

To force the network not to overfit to particular values that are not known (the actual positions of the point sources, $\alpha$, or the specific background model used for example), many images are generated that span the possible space of values. The network has to generalize over these values to achieve reasonable accuracy. This means many realizations of the simulations are needed. As long as the real data is somewhere inside the training box spanned by $f_{\rm src}$, randomized locations of unresolved point source population and the background model, and the ConvNet generalizes over all these values, the network output is reliable.

\end{appendices}

\section*{Acknowledgements}

The {\it Fermi} LAT Collaboration acknowledges generous on-going support from a number of agencies and institutes that have supported both the development and the operation of the LAT as well as scientific data analysis. These include the National Aeronautics and Space Administration and the Department of Energy in the United States; the Commissariat a l'Energie Atomique and the Centre National de la Recherche Scientifique/Institut National de Physique Nucléaire et de Physique des Particules in France; the Agenzia Spaziale Italiana and the Istituto Nazionale di Fisica Nucleare in Italy; the Ministry of Education, Culture, Sports, Science and Technology (MEXT), High Energy Accelerator Research Organization (KEK), and Japan Aerospace Exploration Agency (JAXA) in Japan; and the K. A. Wallenberg Foundation, the Swedish Research Council, and the Swedish National Space Board in Sweden. Additional support for science analysis during the operations phase is gratefully acknowledged from the Istituto Nazionale di Astrofisica in Italy and the Centre National d'Etudes Spatiales in France.

The authors thanks Mattia di Mauro, Manuel Meyer, Gabriela Zaharijas, Christoph Weniger and Tom Heskes for fruitful comments on the manuscript.
R. RdA, is supported by the Ram\'on y Cajal program of the Spanish
MICINN and also thanks the support of the Spanish MICINN's
Consolider-Ingenio 2010 Programme  under the grant MULTIDARK
CSD2209-00064, the Invisibles European ITN project
(FP7-PEOPLE-2011-ITN, PITN-GA-2011-289442-INVISIBLES, the  
``SOM Sabor y origen de la Materia" (FPA2011-29678), the
``Fenomenologia y Cosmologia de la Fisica mas alla del Modelo Estandar
e lmplicaciones Experimentales en la era del LHC" (FPA2010-17747) MEC
projects and the Spanish MINECO Centro de Excelencia Severo Ochoa del IFIC 
program under grant SEV-2014-0398. The work of GAGV was supported by Programa FONDECYT Postdoctorado under grant 3160153.

SC and LH acknowledge the support within the "idark" program of the Netherlands eScience Center (NLeSC).

\bibliographystyle{ieeetr}
\bibliography{bibliography.bib}

\end{document}